\documentclass[preprint,sort&compress,12pt]{elsarticle}

\usepackage{graphicx}
\usepackage{amssymb}
\usepackage{amsthm}
\usepackage{lineno}
\usepackage{fullpage}
\usepackage{mathtools}
\usepackage{booktabs}
\usepackage{multirow}
\usepackage{array}
\usepackage{url}
\usepackage[section]{placeins}
\usepackage{bm}
\usepackage[ruled,linesnumbered]{algorithm2e}
\usepackage{gensymb}
\usepackage{lineno}
\usepackage[colorlinks=true]{hyperref}
\usepackage{subfig}
\usepackage{natbib}

\newcolumntype{P}[1]{>{\centering\arraybackslash}m{#1}}

\journal{Science Bulletin}

\begin{document}

\begin{frontmatter}

\title{Seeing Permeability From Images: \\ 
Fast Prediction with Convolutional Neural Networks}

\author[vt]{Jinlong Wu}
\ead{jinlong@vt.edu}
\author[mines]{Xiaolong Yin\corref{corxh}}
\ead{xyin@mines.edu}
\author[vt]{Heng Xiao\corref{corxh}}
\ead{hengxiao@vt.edu}
\cortext[corxh]{Corresponding author}
\address[vt]{Kevin T. Crofton Department of Aerospace and Ocean Engineering, Virginia Tech, Blacksburg, VA 24060, USA}
\address[mines]{Department of Petroleum Engineering, Colorado School of Mines, Golden, CO 80401, USA}

\begin{abstract}
Fast prediction of permeability directly from images enabled by image recognition neural networks is a novel pore-scale modeling method that has a great potential. This article presents a framework that includes (1) generation of porous media samples, (2) computation of permeability via fluid dynamics simulations, (3) training of convolutional neural networks (CNN) with simulated data, and (4) validations against simulations. Comparison of machine learning results and the ground truths suggests excellent predictive performance across a wide range of porosities and pore geometries, especially for those with dilated pores. Owning to such heterogeneity, the permeability cannot be estimated using the conventional Kozeny--Carman approach. Computational time was reduced by several orders of magnitude compared to fluid dynamic simulations. We found that, by including physical parameters that are known to affect permeability into the neural network, the physics-informed CNN generated better results than regular CNN, however improvements vary with implemented heterogeneity.
\end{abstract}

\end{frontmatter}


\section{Introduction}
In geoscience and engineering, image-based pore-scale studies immediately emerged as abilities to scan high-resolution images of porous rocks became available~\cite{petrovic1982soil,vinegar1987tomographic,flannery1987three}. X-ray computed tomography can be used to construct three-dimensional images of porous rocks with sub-micron resolution up to the scale of 1000$^3$ pixels. Scanning Electron Microscopy (SEM) can reach a resolution of nanometers ($10^{-9}$~m). When combined with Focused Ion Beam (FIB) technology, three-dimensional images of nanometer resolution can be constructed by milling the sample layer by layer~\cite{tomutsa2007analysis}. Image-based analyses have revealed rich pore-scale features previously unavailable, and have become a very useful tool of petrophysics~\cite {wildenschild2002using,arns2005pore,sok2010pore,wildenschild2013x,king2015pore,wu2017mutiscale}.

Computation of pore-scale transport properties from pore-scale images is an important aspect of image-based pore-scale studies. Such computations are generally performed in two ways,  i.e., direct simulation approach and simplified network approach. In the first approach, the microscopic transport equations are solved directly on the geometry shown by the pore-scale images to obtain averaged properties such as permeability, relative permeability, or dispersion coefficient. Both single and multiphase flows can be accounted for, and both reactive and non-reactive transport equations can be solved. This direct approach is generally considered to be more accurate, but the computational cost is very high.  For processes such as multiphase flows and reactive transport with slow kinetics, it is nearly impossible to solve the governing equations in a medium of even a moderate size. Therefore, the second alternative approach is to first abstract the porous medium as a discrete network. By applying simplified flow and transport laws on the network, the computational cost to obtain averaged properties can be effectively lowered~\cite{blunt2013pore}.

Some transport properties of porous media such as permeability are solely functions of pore geometry. Therefore, it should be possible to predict them using a neural network approach, which is to develop a surrogate model that directly maps a pore geometry to physical properties. Such a task resembles that in image classification\cite{lecun1995convolutional,krizhevsky2012imagenet}, where a model takes an image as input and give the classification label as output by recognizing the object in the image, e.g., cars, animals, or even subtypes thereof (i.e., car make or animal breed). Once constructed, such surrogate models can potentially enable fast prediction of physical properties of porous media without performing direct simulations or network calculations. The recent studies of chemical imaging of rocks also involve surrogate models. For example, Hao et al.~\cite{hao2018cross} generated a molecular distribution map across scales by building a machine learning model.

Convolutional neural network (CNN) has achieved significant successes in image classification~\cite{lecun1995convolutional,krizhevsky2012imagenet,lecun2015deep}. Researchers have adopted CNN to solve various problems in science and engineering, e.g.,
solving the quantum many-body problem~\citep{carleo2017solving}, analyzing gravitational lenses in astrophysics~\citep{hezaveh2017fast},
extracting flow features in resolved flow fields~\cite{strofer2018data}, and serving as surrogate model for parameterized partial differential equations~\cite{khoo2017solving}. Recent studies~\cite{mosser2017reconstruction,cang2017microstructure} also demonstrated that porous media can be reconstructed by using generative adversarial network (GAN) or autoencoder, in which CNN is involved to map between the porous media image and the latent space. CNN has also been used to directly predict effective properties of multiphase materials~\cite{yang2018deep,cecen2018material,cang2018improving}. Yang et al.~\cite{yang2018deep} adopted standard CNN to predict elastic homo-genization linkages for 3-D composite material system. Cang et al.~\cite{cang2018improving} also used an existing CNN architecture (i.e., ResNet~\cite{he2016deep}) to predict material properties from microstructures. The preliminary study of Srisutthiyakorn~\cite{srisutthiyakornpermeability} demonstrated the feasibility of predicting permeability directly from rock images by using CNN. The features of connectivity between neighboring pixels were extracted by performing convolution with all possible cross shape templates. Srisutthiyakorn demonstrated that these extracted features lead to better predictive performance than geometric measurements (Minkowski functionals) passed to a regular neural network. Karpatne et al.~\cite{karpatne2017physics} pointed out that data science models can be further improved by leveraging the scientific knowledge. In this article, we use a physics-informed machine learning framework to combine image information and integral quantities (porosity and specific surface area) in the same neural network. We demonstrate that the physics-informed architecture has in general superior predictive performance compared to the conventional CNN, though in some cases we also noted that it is not significantly better than regular CNN. Assessment of the proposed neural network architecture demonstrates that physics-informed CNN predicts permeabilities to 10\% accuracy for synthetic two-dimensional porous media with a wide range of scenarios (porosities, fraction of dilated pores, and similarity levels between training and prediction datasets).

The rest of the paper is organized as follows. The proposed framework for fast prediction of permeability of porous media from images is introduced in Section~2. The prediction results are presented and analyzed in Section~3. In Section~4, we provide some insights on how CNN predicts permeabilities with high accuracy from the images. Section~5 concludes the paper.

\label{sec:intro}
\section{Methodology}
\label{sec:method}
\subsection{Overview of the computational framework}

\begin{figure*}[!htbp]
  \centering
  \includegraphics[width=0.9\textwidth]{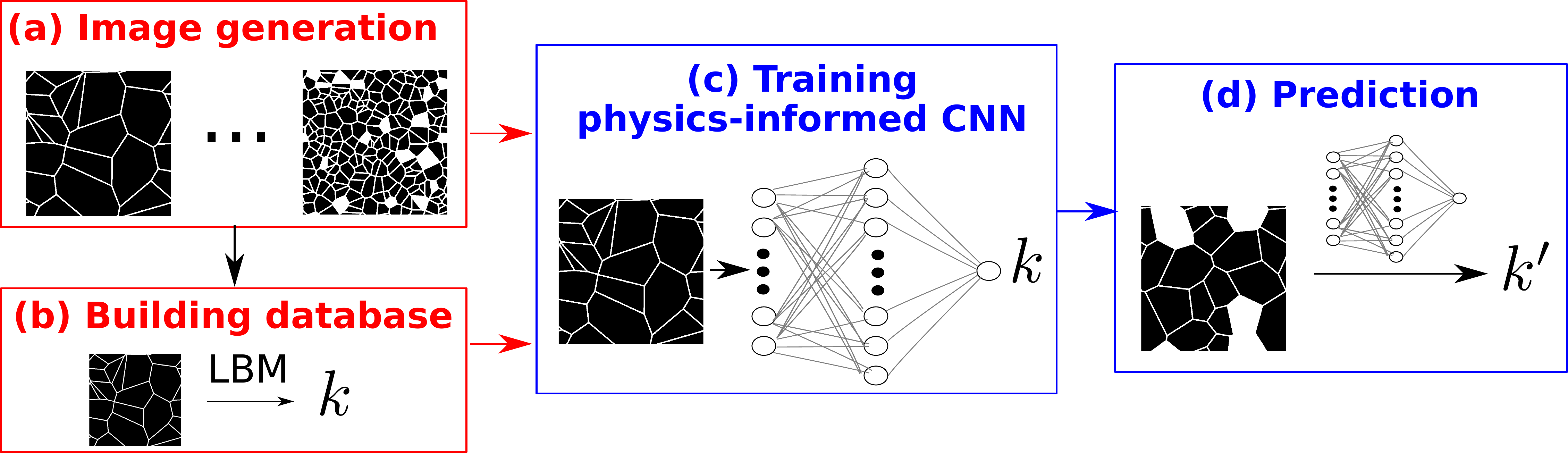}
    \caption{Overview of the framework, including (a) generating images, (b) building a database by performing lattice Boltzmann simulations to obtain permeability, (c) using the database to train physics-informed CNN and (d) predicting the permeability of new samples.}
  \label{fig:overview}
\end{figure*}

\begin{figure}[!htbp]
  \centering
  \includegraphics[width=0.45\textwidth]{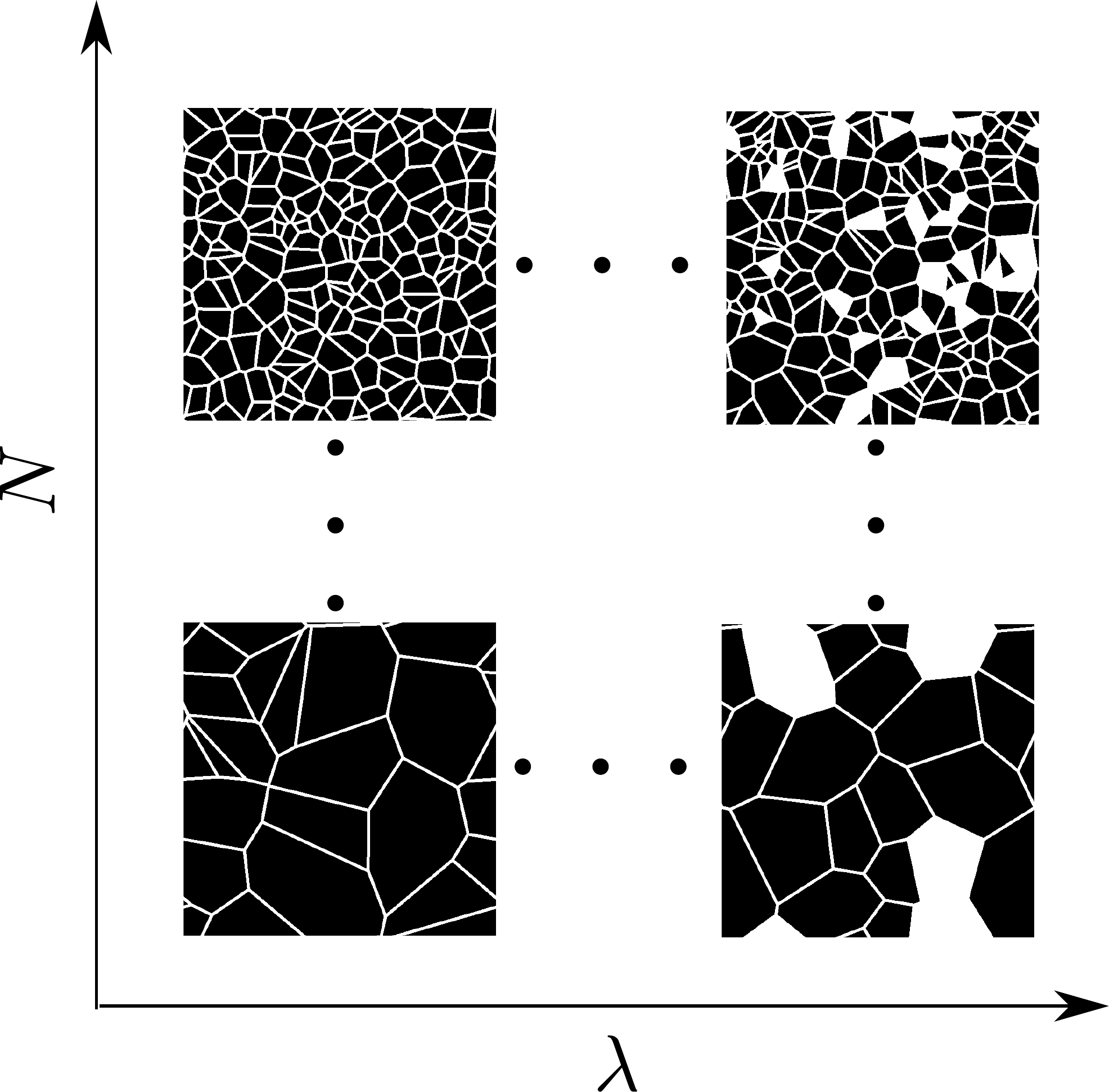}
    \caption{Sample images of porous media in the parameter space. The two algorithmic parameters are $N$ that controls the density of polygonal grains and the porosity and probability $\lambda$ that controls the porosity of dilated pores.}
  \label{fig:sample-rock}
\end{figure}

The objective of the computational framework presented in this study is to train a machine-learning model for fast prediction of permeability. This framework consists of the following steps as illustrated in Fig.~\ref{fig:overview}:
\begin{enumerate}[(i)]
    \item \emph{Generating training dataset.} We first generated a number of images of synthetic porous media covering a wide range of the chosen parameter space (porosity and percentage of dilated pores, see Fig.~\ref{fig:sample-rock}). Direct simulations with lattice Boltzmann method were then used to compute the permeabilities of the generated porous media samples. The image--permeability pairs form the training database for the neural network based machine learning model. The detailed procedure of generating training database is presented in Section~\ref{sec:method-image} and Figs.~\ref{fig:overview}(a)--(b). 
    \item \emph{Training physics-informed CNN model.} The data obtained from the previous step were then used to train a neural network that takes both an image and its physical geometric property (porosity and surface area ratio) as input and gives permeability as output. 
    Details of the CNN architecture and the training procedure are presented in Section~\ref{sec:method-cnn} and Fig.~\ref{fig:overview}(c).
    \item \emph{Predicting permeabilities for new images.} The trained model obtained from step (ii) was then used to provide permeability for new images that are not in the training database. See Fig.~\ref{fig:overview}(d).
\end{enumerate}

Potentially both synthetic and real porous media images can be used when generating the training database in step (i). Regardless of the source of images, it is essential to ensure that samples in the database cover sufficient regions of the parameter space. This requirement is easier to satisfy with synthetic images. In this study, we used two-dimensional synthetic binary images. Three-dimensional, real rock images will be the objective of future studies.

\subsection{Generation of Training Data}
\label{sec:method-image}
\label{sec:method-lbm}

Images of synthetic porous media were generated using a Voronoi tessellation algorithm that has been presented in several earlier studies~\cite{wu2012single,newman2013lattice,yong2014direct}. Voronoi tessellation is a method to partition a plane using a given set of points henceforth mentioned as the \emph{seeds}. Each partition, a Voronoi cell, is a polygon that represents the set of points on the plane that are closer to the enclosed seed than to any other. In our algorithm, seeds were generated randomly in a $600 \times 600$ (pixel) area. The edges of generated Voronoi cells were then given a width of six pixels to form a fully percolated network of flow channels. When channels are six-pixel wide, the permeabilities from lattice Boltzmann simulations are within 5\% of those extrapolated to infinite lattice resolution. 
The number of seeds $N$, or equivalently the number of initial Voronoi cells, in a given domain size is an algorithmic parameter to control the size of polygons in the image and the porosity. The porosity of synthetic media obtained using this algorithm increases from 0.084 $\pm$ 0.002 when $N$ is 18, to $0.257 \pm 0.002$ when $N = 189$. The porosity--permeability relation of these synthetic geometries can be well fitted by the Kozeny--Carman equation:
\begin{equation}
\label{eq:kc}
k = \frac{0.14 \, \phi^3}{s^2 (1-\phi)^2} ,
\end{equation}
where $\phi$ is the porosity and $s$ is the ratio between the total perimeter of the polygons and the total area of the polygons. The coefficient $0.14$ in the numerator is a fitting parameter established by the dataset presented in this work. Note that $s$ is the two-dimensional analogue of specific surface area, the net surface area over the net solid volume and an important characteristics of porous media.

\begin{figure}[!htbp]
  \centering
    \includegraphics[width=0.4\textwidth]{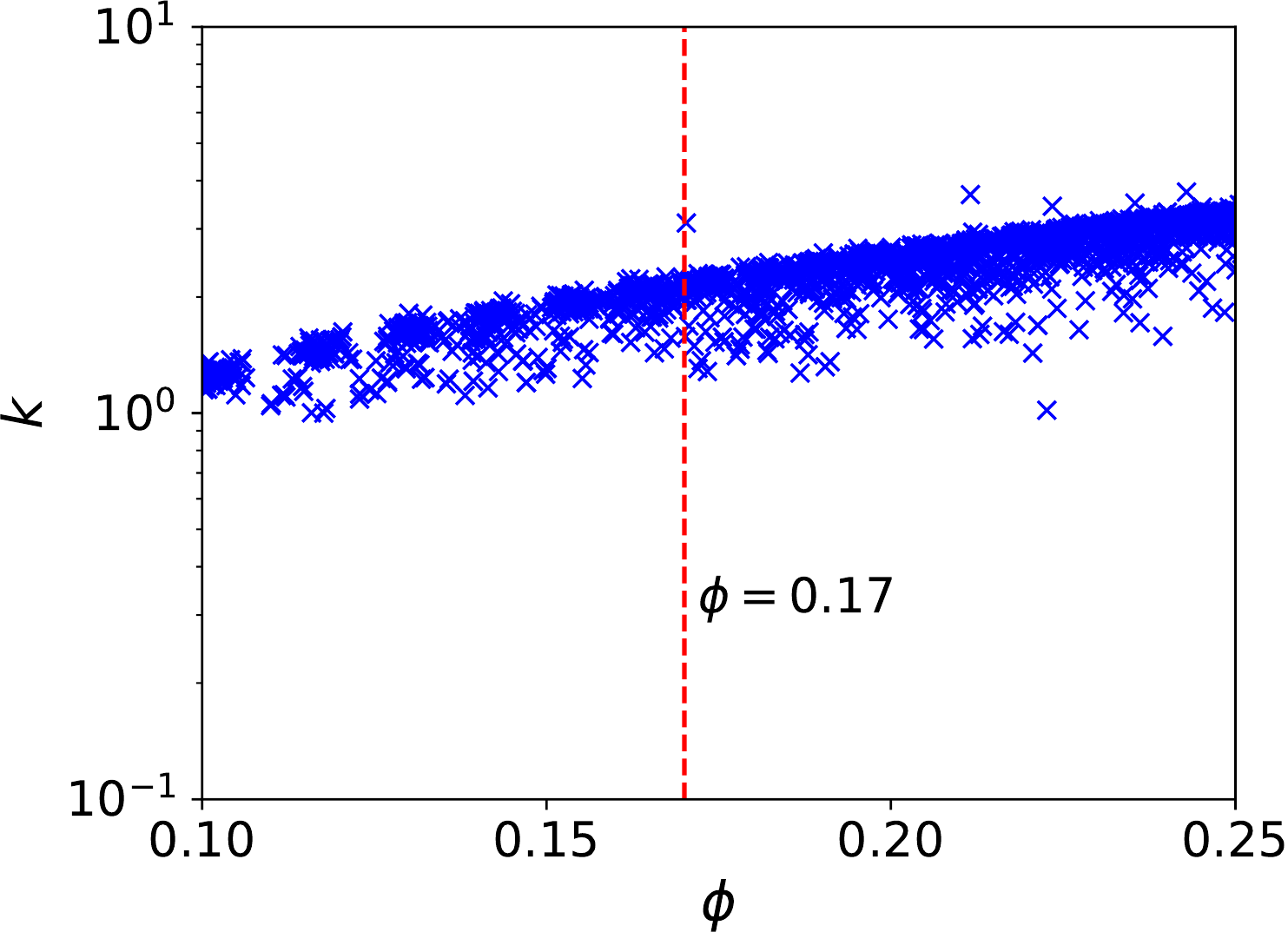}
    \caption{Relation between the permeability and the porosity from our synthetic 2-D models. Permeability is scaled by $3\mu\text{m}/\text{pixel}$ resolution. For porosity greater than 0.17 this relation agrees well with that of Fontainebleau sandstones from Arns et al.~\cite{ams2004virtual}.}
  \label{fig:kappa-phi}
\end{figure}

To introduce more variability in the synthetic geometries, we used a probability $\lambda$ to remove Voronoi cells from generated tessellations. Areas occupied by removed cells were assigned to the fluid, creating large and isolated space that resembles dilated pores found in many geological porous media. 
As such, the algorithmic parameter $\lambda$ is an approximate proxy that controls the porosity of dilated pores in the porous medium. Increasing $\lambda$ effectively increases the porosity of medium while keeping the specific surface area $s$ nearly unchanged, leading to scattering of the $\phi$--$k$ relation. The permeabilities of cases with $\lambda > 0$ cannot be well predicted by the Kozeny--Carman equation established for the case of $\lambda=0$. These cases ($\lambda > 0$) are therefore particularly interesting. In this study, two values of $\lambda$ were used to generate synthetic samples with dilated pores: $\lambda=0.05$ and $0.10$. The porosity of synthetic geometries ($N \in [18, 189]$ and $\lambda \in \{0,\ 0.05,\ 0.10\}$) ranges from $0.08$ to $0.39$, covering typical porosity values of real rocks. The relation between the permeability and the porosity is presented in Fig.~\ref{fig:kappa-phi} for the current database of synthetic geometries. This relation when scaled by a common CT-scan resolution of $3\mu\text{m}/\text{pixel}$ is in good agreement with numerical and experimental data of Fontainebleau sandstone presented in~\cite{ams2004virtual} when the porosity is greater than about $17\%$. When porosity is less than $17\%$, our synthetic geometries have higher permeabilities than~\cite{ams2004virtual}, perhaps due to cementation or blockage of pore throats that are not considered in our current geometries. It should be noted that the purpose of this work is to demonstrate the feasibility of using convolutional neural networks in fast prediction of permeability for porous media and hence we do not seek to obtain exact representations of rock geometries. For this purpose, two different parameters, $N$ and $\lambda$, were introduced to represent a selected complexity of 2-D porous media. Incorporating more complexities in the training is possible, e.g., varying the channel width or even closing some channels. These complexities may be useful for achieving better representations of real-rock geometries.

Permeabilities of the generated synthetic geometries were obtained by using lattice Boltzmann simulations.  Specifically, a two-dimensional, nine-velocity (D2Q9) scheme with the multi-relaxation time collision operator that we wrote and presented in an earlier paper~\cite{newman2013lattice} was used was used in the simulations. Fluid flow from one side of the image to the opposite side was generated by a body force or acceleration assigned to the fluid. The body force and the viscosity of the fluid were chosen such that flows were strictly in the Stokes regime. Domain-averaged, steady-state velocity of the fluid was used to compute the permeability. Details of the lattice Boltzmann method can be found in our supporting information, and critical methods and parameters are summarized in Table~\ref{tab:lbm-summary}.  All lengths in this study have the dimension of pixel, which is the native unit of digital images. Consequently, permeabilities of all synthetic geometries are reported in the unit of pixel squared. Permeability expressed in pixel squared can be related to the dimensional permeability by the resolution of the image. For instance, if the resolution of an image is 2.0 $\mu\textrm{m}$ per pixel and the permeability of the image is 0.25 $\textrm{pixel}^2$, the permeability of the medium would be $2.0^2 \times 0.25 = 1.0$ $\mu\textrm{m}^2$. Previously presented Fig.~\ref{fig:kappa-phi} is an example of this conversion. The computational domain size of lattice Boltzmann simulations is  $600\times600$ pixels, and the computational time for one simulation is approximately 1 hour on a single CPU core.
\renewcommand{\arraystretch}{1.1}
\begin{table}[htbp] 	
  \centering
  \caption{Settings of lattice Boltzmann simulations.}
\label{tab:lbm-summary}
\begin{tabular}{P{1.5cm}  P{1.5cm}  P{2.5cm} P{6.0cm}  P{2.0cm} }	
\toprule
  Scheme & Collision & Boundary condition & Body force (Pressure gradient equivalent ) & Kinematic viscosity \\ 
 \hline
 D2Q9 & MRT & Bounce-back & $\Delta P/\rho L=2.78\times10^{-9} \Delta x/\Delta t^2$ & $1/6 \Delta x^2/\Delta t$ \\
\bottomrule
\end{tabular}
\end{table}

\subsection{Convolutional Neural Network}
\label{sec:method-cnn}

Neural networks are a class of machine learning models that are parameterized by coefficient vector ${W}$ and represent mappings from input~$\mathbf{q}$ to output~$\bm{y}$ in the form of a sequence of composite functions. For example, a neural network with one layer of intermediate variables between input and output (one \emph{hidden} layer) may be represented by the following composite functional mapping:
 \[\bm{y} = {W}^{(2)} \sigma \left({W}^{(1)} \mathbf{q} + \mathbf{b}^{(1)}\right) + \mathbf{b}^{(2)},\] 
 or alternatively written in an equivalent form as:
 \[
 \bm{y} = {W}^{(2)} \mathbf{h} + \mathbf{b}^{(2)}
 \qquad \text{with} \qquad
 \mathbf{h} = \sigma \left({W}^{(1)} \mathbf{q} + \mathbf{b}^{(1)}\right) ,
 \]
where $\sigma$ is an \emph{activation function} such as $\sigma(q) = \tanh(q)$ or $\sigma(q) = 1/(1+e^{-q})$; ${W}^{(i)}$ and $\mathbf{b}^{(i)}$ indicate weights and bias, respectively, of the $i^{\text{th}}$ layer. In the context of this work, the input $\mathbf{q} \in \mathbb{R}^{600 \times 600}$ is the binary image of $600 \times 600$ pixels, and the output $\bm{y} \in \mathbb{R}$ is the permeability. 

Compared to the fully connected neural networks, convolution neural networks (CNN) exploit two facts to significantly reduce the number of coefficients and thus to increase learning efficiency. First, a neuron, the basic unit of a neural network, is only locally connected to several neurons in the previous layer as spatially nearby pixels in an image are more correlated. Second, the output sought from images has translational invariance ~\cite{bishop2006pattern}, which allows weight sharing of convolution kernel at all locations. Such preservation of invariance allows CNN to achieve a comparable performance of regular neural networks with much less training data.

\begin{figure*}[!htbp]
  \centering
  \includegraphics[width=0.98\textwidth]{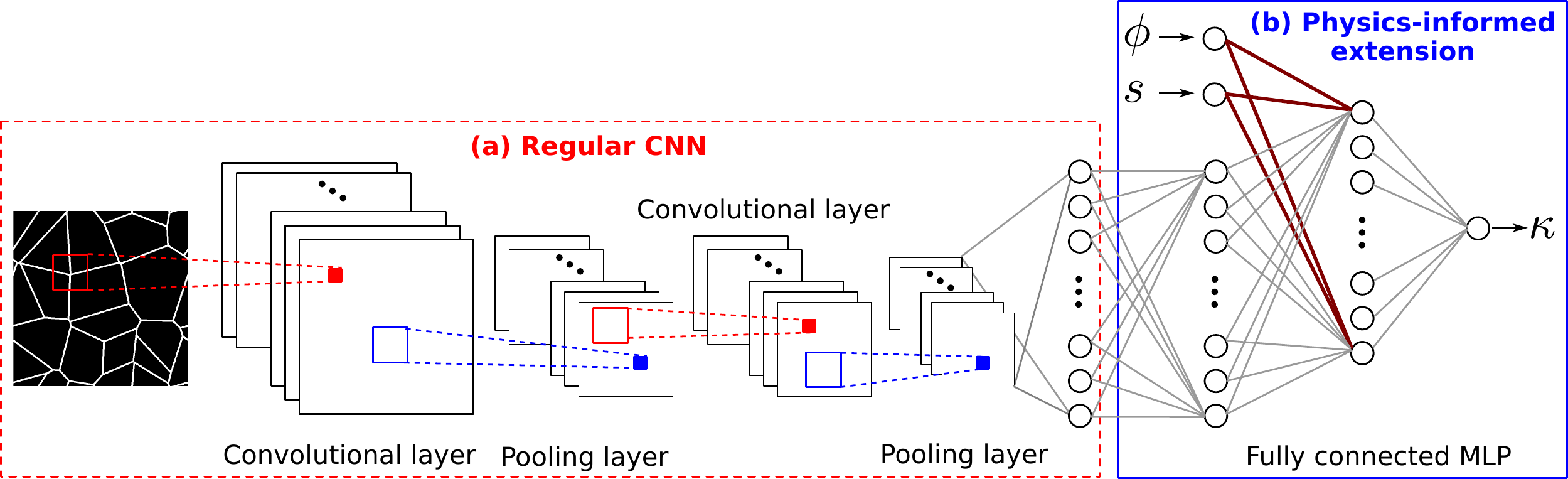}
    \caption{Physics-informed CNN architecture, including (a) regular CNN and (b) physics-informed CNN by introducing a fully connected multilayer perceptron (MLP) neural network. In this work, the input image has been preprocessed by a $6\times6$ max pooling kernel to reduce the image size and thus the training cost.}
  \label{fig:PI-CNN}
\end{figure*}

A CNN consists of a number of convolutional layers and pooling layers, followed by fully connected layers. In the problem of estimating permeability from images as concerned in this study, the convolutional and pooling layers mapped the image space to physical quantity space. The fully connected layers represent a nonlinear mapping between physical quantities, with permeability as the final output. We extended the regular, image-classification CNN architecture (Fig.~\ref{fig:PI-CNN}a) by introducing the porosity $\phi$ and the specific surface area $s$ into one of the fully connected layers (see the thick/red edges in Fig.~\ref{fig:PI-CNN}b). The extended network architecture is referred to as \emph{physics-informed CNN} in view of its relation with our previous work that used machine learning for physical modeling~\cite{wang2017physics-informed,wu2018physics}. Both $\phi$ and $s$ are parameters of the Kozeny--Carman equation. Their influence on the permeability $k$ was built into the neural network architecture in an explicit yet flexible manner. That is, the proposed network architecture represents our prior knowledge that the permeability $k$ \emph{may be} a function of $\phi$ and $s$, but the specific functional relation is not known and needs to be established by training. While $\phi$ and $s$ were the most natural choices due to their connections to the Kozeny--Camen equation, other choices are possible but not tested in this study. The CNN architecture proposed here is inspired by earlier works in image classification~\cite{lecun1995convolutional,krizhevsky2012imagenet,lecun2015deep} and physical modeling~\cite{karpatne2017physics}. The proposed CNN architecture is implemented in machine learning frameworks \verb+Lasagne+ and \verb+Theano+~\cite{lasagne,theano}.

The CNN used in our study includes two convolutional layers, each followed by a pooling layer, which are then followed by three consecutive fully-connected layers. Each convolutional layer has 10 channels and a convolutional kernel of size $5\times5$ to extract different features from the corresponding input. In the two pooling layers, the max pooling function and a kernel size of $2\times2$ were adopted. The three fully-connected layers have 10, 32, and 10 neurons, respectively. Among the neurons in the second fully connected layer are two neurons representing porosity $\phi$ and specific surface area $s$. The number of layers was empirically selected to ensure enough complexity of the neural network. The number of channels or neurons within each layer were chosen by using grid searching to minimize the mean squared error of the predicted permeabilities for the training database. Instead of using the cross-validation, dropout ratios of 0.2 and 0.1 were applied to the first and the third fully-connected layers, respectively, to avoid the overfitting of the trained model~\cite{srivastava2014dropout}. The computational cost is of the order of seconds for predicting the permeability of a $600\times600$ pixels image by using the trained physics-informed CNN in Fig.~\ref{fig:PI-CNN}. This is three orders of magnitude lower than lattice Boltzmann simulations. Note that what is not shown in Fig. 3 is a $6\times6$ max pooling kernel used to preprocess the images before inputting the images to CNN. The purpose of this procedure is to reduce the image size and the computational cost of CNN training.

The architecture should also be applicable to 3-D porous media, for which we need to use a 3-D kernel for the convolutional neural network. Therefore, the convolutional neural network will have more coefficients and the training cost will increase accordingly. In addition, representing the 3-D porous media in pixels leads to a higher dimensional space, which will require more training data and thus more training computational cost.

\section{Numerical Experiments}
\label{sec:results}

In this section we present results of three numerical experiments. We first show the merit of the proposed framework by demonstrating the predictive performance in cases with no dilated pores for a wide range of the number of seeds $N$ (case~1). Further, in case~2 we show that the neural network model trained by a diverse dataset consisting of samples with and without dilated pores ($\lambda=0$ and 0.1, respectively) is able to predict samples with pore heterogeneity not in the training dataset ($\lambda = 0.05$). Finally, in case~3 we show that neural network model has a significantly better predictive performance than the Kozeny--Carman equation for rocks with dilated pores ($\lambda=0.05$). The porosity distribution of the prediction set for case 3 is $0.233\pm 0.002$. The prediction set for case 3 is chosen such that the number of seeds $N=108$ is in the middle of the range of the number of seeds from the training set,  but not within the training set. Detailed setup and parameters of the three cases including the number of images $n$, number of seeds $N$, and $\lambda$ for both training and prediction datasets are presented in Table~\ref{tab:cases}.

\begin{table}[htbp] 	
  \centering
  \caption{Parameters of cases investigated in this work including the number of images in dataset $n$, the number of seeds $N$, and $\lambda$ for both training and prediction. Square and curly brackets are used to indicate range (with intervals) and sets, respectively. For example, the range/interval notation $[1 : 2 : 9]$ is equivalent to set $\{1, 3, 5, 7, 9\}$.}
\label{tab:cases}
\begin{tabular}{c|ccc|ccc}  
\toprule
 \multirow{2}{*}{case no.} & \multicolumn{3}{c|}{training set} & \multicolumn{3}{c}{prediction set} \\ 
{} & $n$ & $N$ & $\lambda$ & $n$ & $N$ & $\lambda$ \\
 \hline
 1 & 980 & [18 : 9 : 189] & 0 & 20 & [18 : 9 : 189] & 0 \\
 2 & 1960 & [18 : 9 : 189] & \{0, 0.1\} & 20 & [18 : 9 : 189] & 0.05 \\
 3 & 490 & [27 : 18 : 189] & 0.05 & 50 & 108 & 0.05 \\
\bottomrule
\end{tabular}
\end{table}

\begin{figure*}[!htbp]
  \centering
  \subfloat[]{\includegraphics[width=0.49\textwidth]{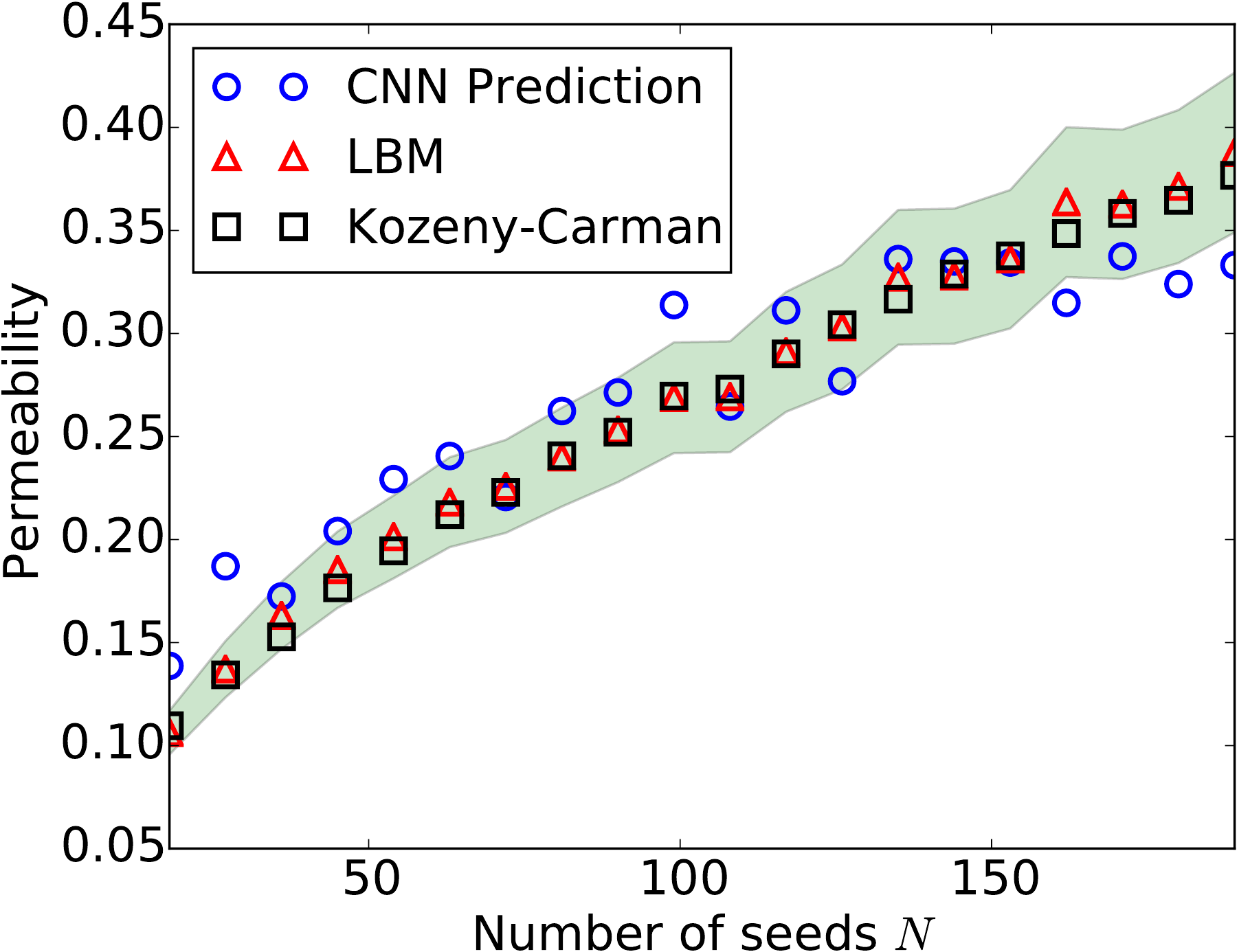}}\hspace{0.3em}
  \subfloat[]{\includegraphics[width=0.49\textwidth]{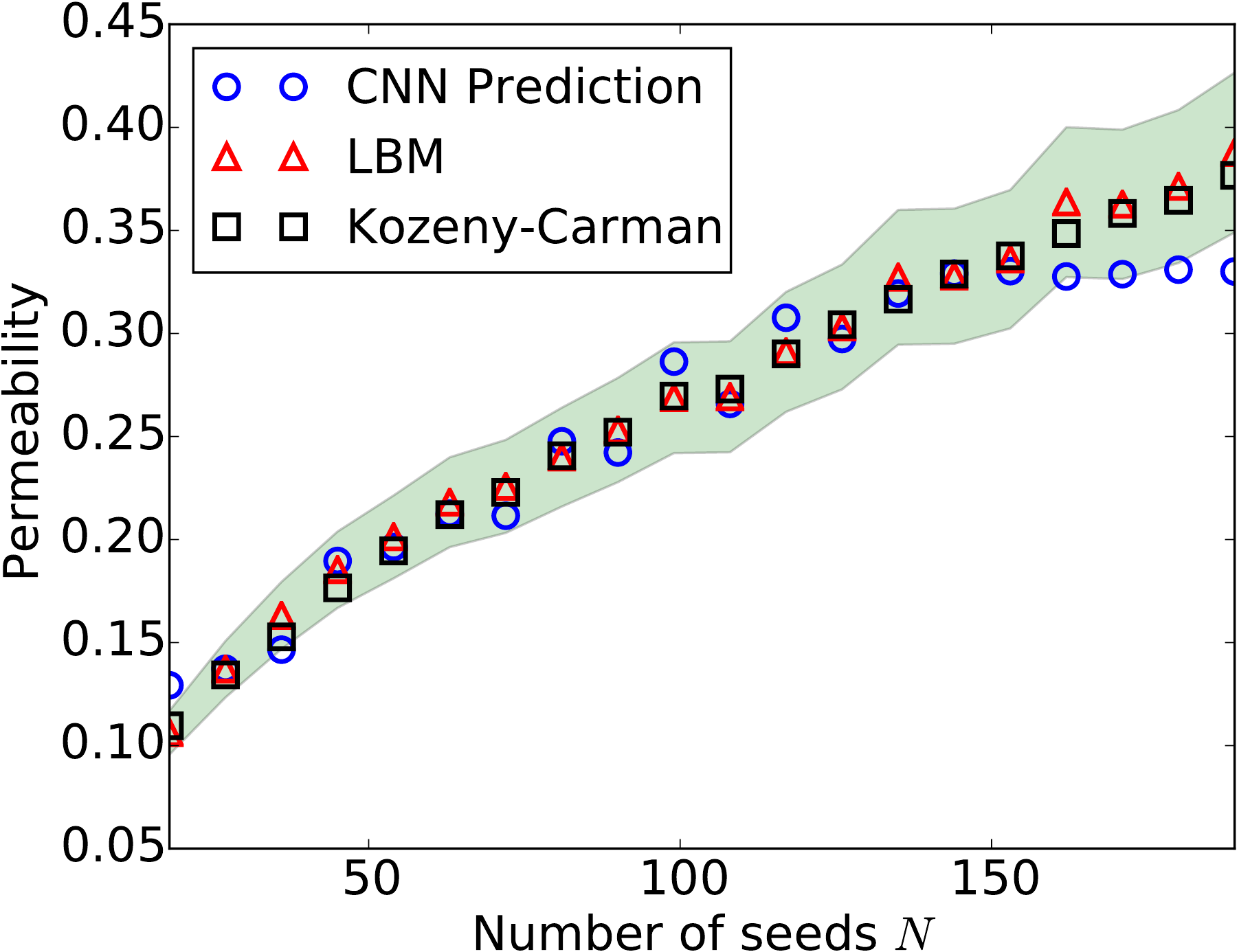}}
    \caption{Prediction of permeability for case~1 by using (a) the regular CNN and (b) the physics-informed CNN.}
  \label{fig:perc1}
\end{figure*}

The regular CNN provides satisfactory predictions of permeabilities as demonstrated in case 1, with most data points falling within an error range of $\pm 10\%$ (shown as shaded regions in Fig.~\ref{fig:perc1}a).   Here, the Lattice Boltzmann simulation results are taken as ground truth, since training data were provided by such simulations. The Kozeny--Carman equation clearly has a better performance than the regular CNN model, which is expected as the rock samples in this case do not have dilated pores. However, by incorporating the physical quantities $\phi$ and $s$ into the network architecture, predictions from the physics-informed CNN showed significant improvements. For most data points the physics-informed CNN predicts permeabilities very close to those from the Kozeny--Carman equation  (see Fig.~\ref{fig:perc1}b). This comparison clearly shows the superiority of the physics-informed CNN to regular CNN. Underestimation of permeability can be observed when the number of seeds $N>150$ in Fig.~\ref{fig:perc1}. The main reason is that the preprocessing procedure with a $6\times6$ max pooling kernel reduces the resolution of the original images. Specifically, the CNN had difficulty in distinguishing the images with different number of seeds $N$ when $N$ is large. Therefore, the CNN tends to underestimate the permeability when predicting the cases with number of seeds $N>150$.

\begin{figure}[htbp]
  \centering
  \subfloat[]{\includegraphics[width=0.49\textwidth]{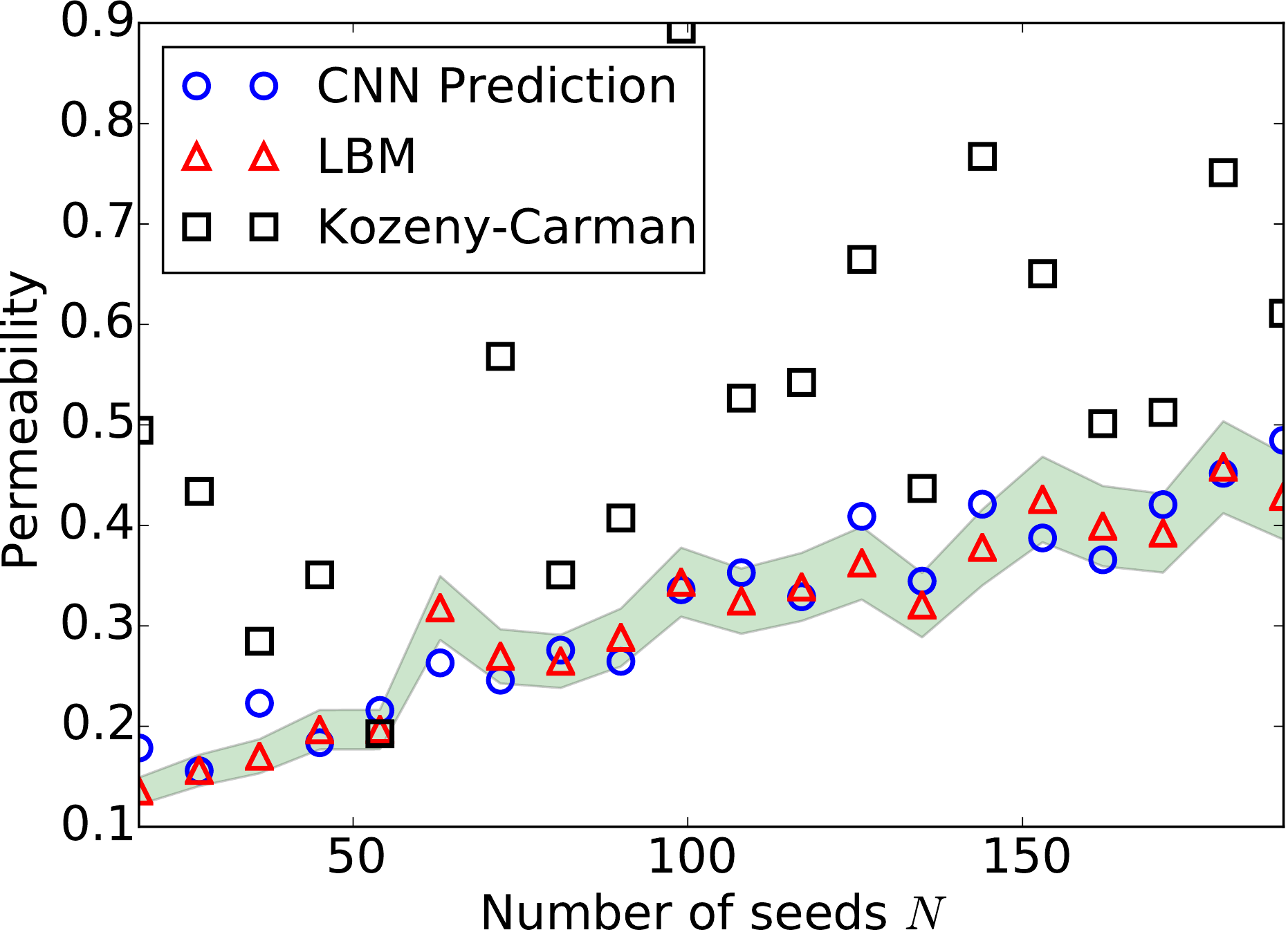}}\hspace{0.3em}
  \subfloat[]{\includegraphics[width=0.49\textwidth]{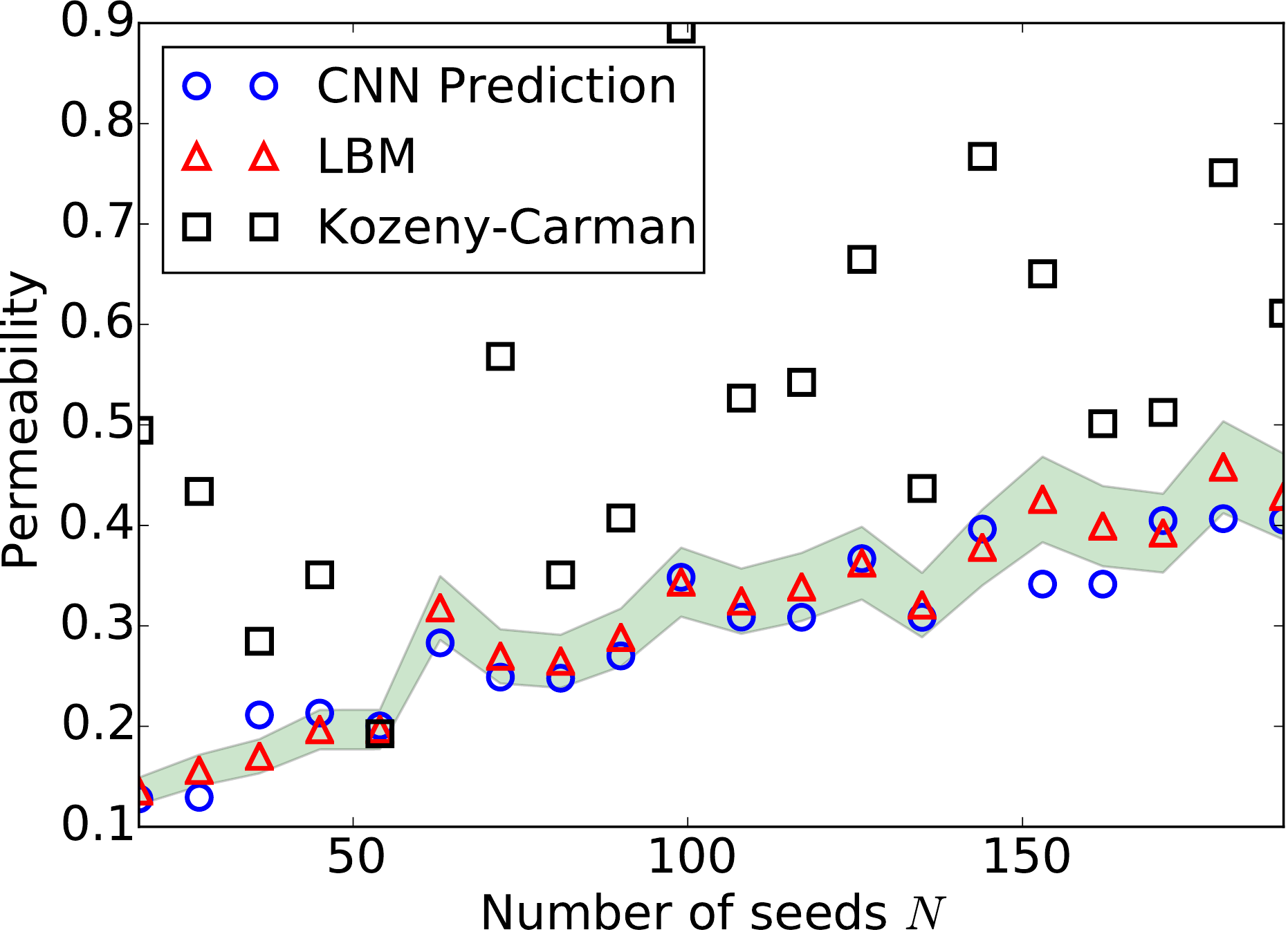}}
\caption{Prediction of permeability for case~2 by using (a) the regular CNN and (b) the physics-informed CNN.}
  \label{fig:perc0p95}
\end{figure}

Case 2 demonstrates that the physics-informed CNN is able to predict permeabilities within $\pm 10\%$ error range of the ground truth for samples with dilated pores (shown in Fig.~\ref{fig:perc0p95}b), even though the training datasets have different $\lambda$. In comparison, the Kozeny--Carman equation is not able to give accurate predictions, with relative errors over 200\%. The physics-informed CNN has the capability of exploring more accurate functional mappings from the training database by taking into account information from the entire image. In contrast, the Kozeny--Carman equation only has  physical variables $\phi$ and $s$, and thus is not able to account for the presence of dilated pores. For this case, the neural network model is clearly more flexible in formulating the functional mapping compared to the analytical formula. This is ultimately attributed to the capability of neural networks in representing high-dimensional mapping (from $\mathbb{R}^{600 \times 600}$ to $\mathbb{R}$), allowing it to take the entire image as input. As explained in Section~\ref{sec:method-cnn}, although the physics-informed CNN contains variables $\phi$ and $s$ as neurons, it may not utilize them when they do not contribute in explaining the permeability data (e.g., in the presence of dilated pores). Therefore, the predicative performance of the physics-informed CNN for this case is similar to results of the regular CNN as shown in Fig.~\ref{fig:perc0p95}. However, the similar performance between the physics-informed CNN and the regular CNN for this case does not mean that the physical neurons $\phi$ and $s$ are not needed. The impact of a given neuron upon the neural network output can be analyzed by studying the trained weights of the neural network. Based on the analysis of the weights in physics-informed CNN, the impact of physical neurons $\phi$ and $s$ on the permeability prediction becomes smaller, but still significant, when the probability $\lambda$ is larger than zero and dilated pores exist.

\begin{figure*}[!htbp]
  \centering
  \subfloat[]{\includegraphics[width=0.49\textwidth]{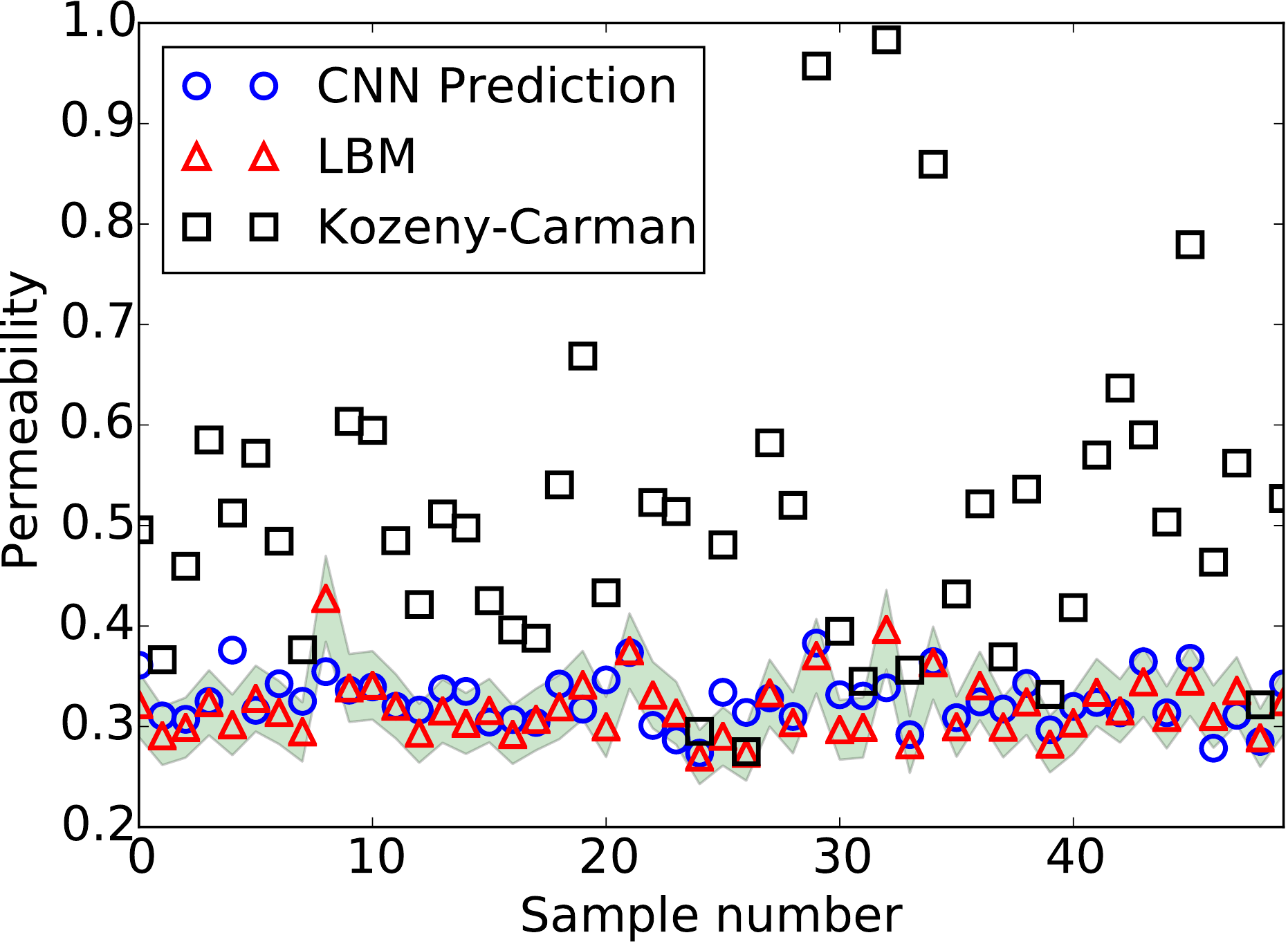}}\hspace{0.3em}
  \subfloat[]{\includegraphics[width=0.49\textwidth]{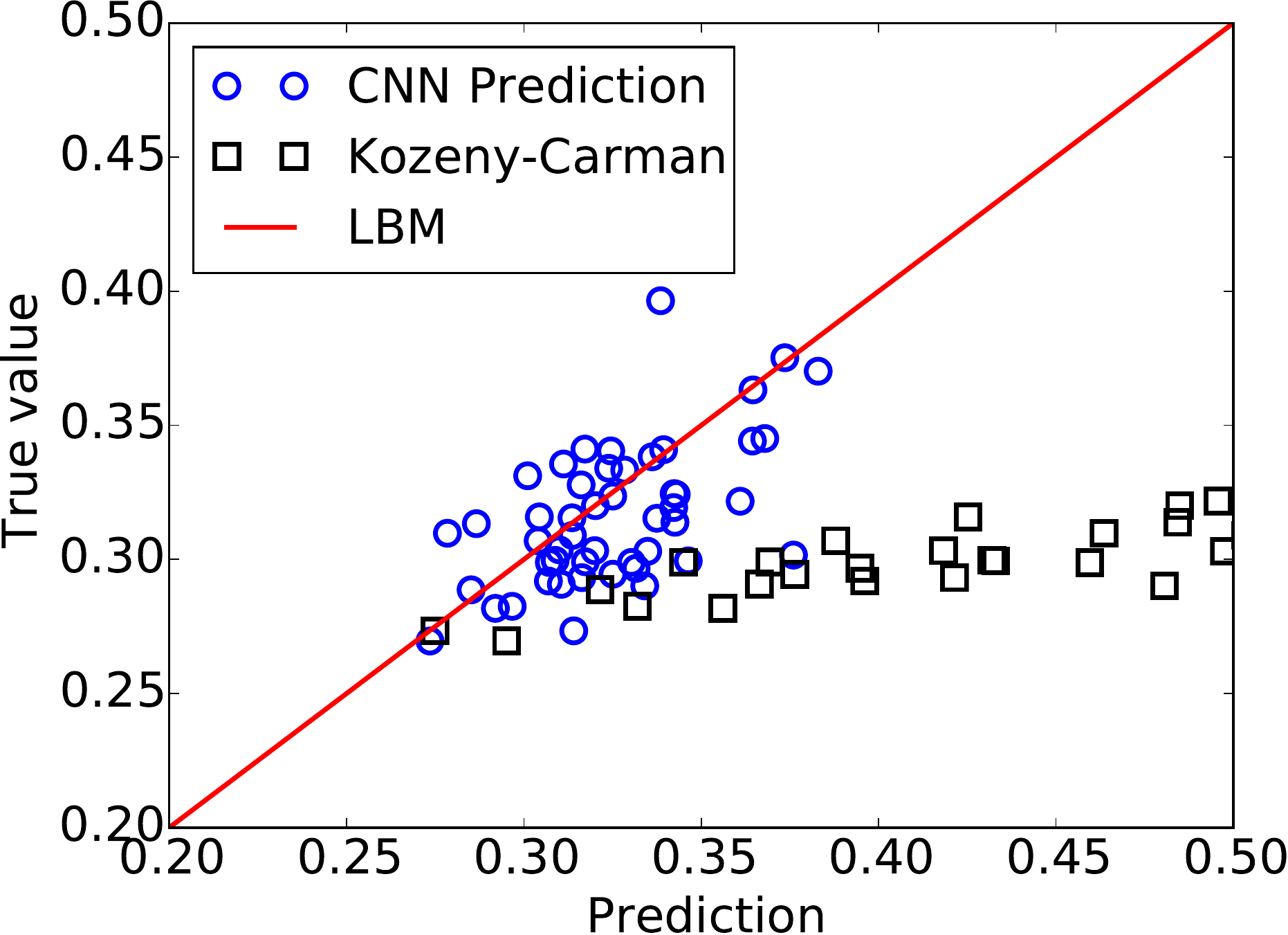}}
    \caption{Prediction of permeability based on physics-informed CNN for case~3, including (a) the permeability against the sample index and (b) the comparison of the prediction and the ground truth.}
  \label{fig:N-pred}
\end{figure*}

In case 3 we further highlight the predictive capability of the physics-informed CNN by using training and testing datasets with different $N$. It can be seen in Fig.~\ref{fig:N-pred}a that the Kozeny-Carman equation again overestimates the permeability of all testing samples due to the presence of dilated pores. The predictions of the physics-informed CNN show much better agreement with ground truth (lattice Boltzmann simulations) than the Kozeny--Carman and mostly fall within the $\pm 10\%$ error range. The improvement of CNN prediction over the Kozeny--Carman equation can be clearly seen by plotting the predictions against the ground truth in Fig.~\ref{fig:N-pred}b, where CNN predictions align much better with the ground truth indicated by the solid line. It should be noted that the view in Fig.~\ref{fig:N-pred}b is zoomed to better present the prediction of physics-informed CNN. Some results based on Kozeny--Carman equation are significantly different from the ground truth and thus cannot be seen in this zoom-in view.

$\text{R}^2$ scores and mean squared errors are shown in Table~\ref{tab:error-summary} to provide more quantitative evaluation of the results. It can be seen that the machine learning predictions are much better than Kozeny--Carman in cases 2 and 3, where dilated pores exist. Compared to the standard CNN, the physics-informed CNN has better prediction performance in all three cases, though the improvement in case 2 is relatively marginal. The definition of $\text{R}^2$ scores and mean squared errors are presented as follow:
\begin{equation}
R^2=1-\frac{\sum\limits_{i} (\kappa_i^{\text{CNN}}-\kappa_i^{\text{LBM}})^2}{\sum\limits_{i} (\kappa_i^{\text{LBM}}-\overline{\kappa^{\text{LBM}}})^2}
\end{equation}

\begin{equation}
\text{MSE}=\frac{1}{n}\sum\limits_{i=1}^{n} (\kappa_i^{\text{CNN}}-\kappa_i^{\text{LBM}})^2
\end{equation}
where the superscript CNN and LBM denote the results from CNN and lattice Boltzmann simulation, respectively, and the term $\overline{\kappa^{\text{LBM}}}$ represents the average of the lattice Boltzmann simulation results. Note that $R^2$ scores of case 3 are not as good as those for cases 1 and 2, suggesting that, despite of the visual agreement seen in Fig.~\ref{fig:N-pred}, case 3 is a more difficult case where the differences between the predictions and the ground truth become as significant as the variations in the ground truth. On case 3, physics-informed CNN shows more improvement over CNN than case 2.

\begin{table}[htbp] 	
  \centering
  \caption{$\text{R}^2$ scores and mean squared errors of the results by using Kozeny--Carman (K-C) equation, standard CNN and physics-informed CNN (PI-CNN).}
\label{tab:error-summary}
\begin{tabular}{c|ccc|ccc}  
\toprule
 \multirow{2}{*}{case no.} & \multicolumn{3}{c|}{$\text{R}^2$ score} & \multicolumn{3}{c}{Mean squared error} \\ 
{} & K-C & CNN& PI-CNN & K-C & CNN & PI-CNN \\
 \hline
 1 & 0.993338 & 0.861430 & 0.926315 & 0.000042 & 0.000883	 & 0.000470 \\
 2 & $-7.361731$ & 0.878642 & 0.884680 & 0.074112 & 0.001076 & 0.001022 \\
 3 & $-65.155383$ & $-0.714258$ & 0.204947 & 0.059938 & 0.001553 & 0.000720 \\
\bottomrule
\end{tabular}
\end{table}

\section{Discussion}
\label{sec:discussion}
It is important to understand why CNN provided such good predictions of permeabilities with a limited set of training data.  A critical prerequisite of good prediction is that the permeability is indeed a function of the pore geometry, and thus a functional mapping from the rock pore geometry to permeability is expected to exist. Attempting to fit a functional relation that does not exist would fail regardless of how sophisticated the machine learning model is. We provide below some insights on how CNN predicated permeability from images pixels information.

\begin{figure*}[!htbp]
  \centering
  \subfloat[]{\includegraphics[width=0.3\textwidth]{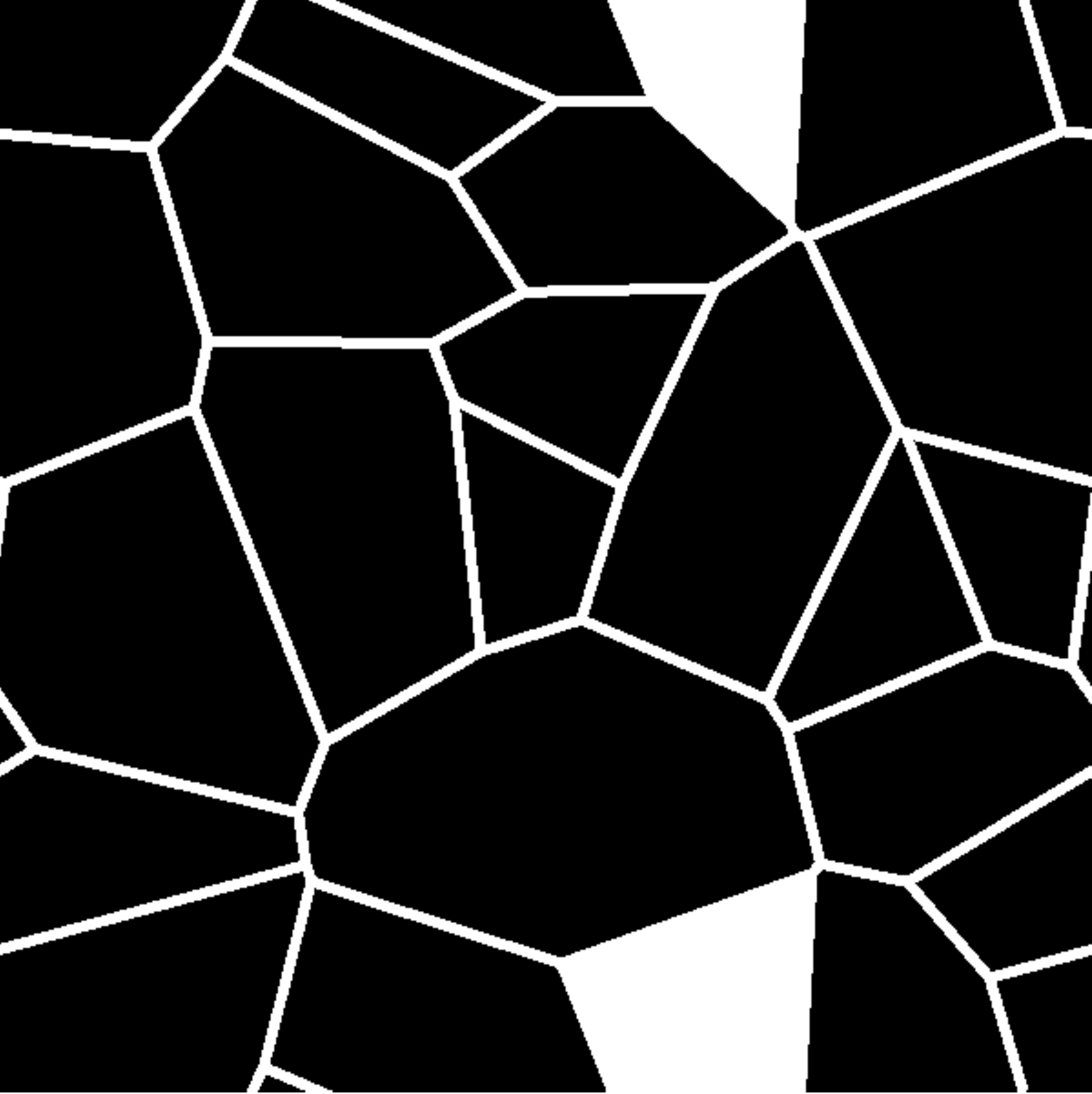}}\hspace{0.3em}
  \subfloat[]{\includegraphics[width=0.3\textwidth]{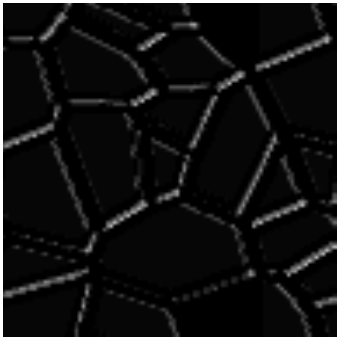}}\hspace{0.3em}
  \subfloat[]{\includegraphics[width=0.3\textwidth]{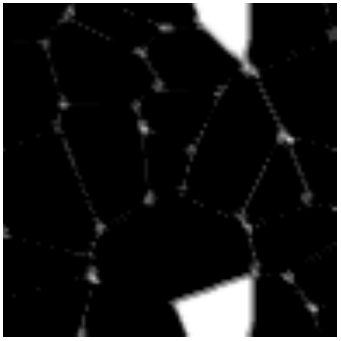}}
    \caption{Illustration of features learned by CNN, including (a) an input image and (b,c) two corresponding features obtained from the first convolutional layer.}
  \label{fig:features}
\end{figure*}

In a CNN each convolutional layer contains the filtered results of the previous layer, and these filtered results can be visualized to illustrate what the CNN learns. Here we use filtered results from the first convolutional layer to show how the trained CNN analyzes an unseen image and makes the corresponding prediction. Two typical filtered results from the first convolutional layer are presented in Figs.~\ref{fig:features}b and c for the prediction of permeability of a rock sample image shown in Fig.~\ref{fig:features}a. It can be seen that in Fig.~\ref{fig:features}b CNN attempts to identify all the paths and temporarily ignores the dilated pores (removed Voronoi cells). On the other hand, the removed Voronoi cells are treated together with the connection points by another filtered result as shown in Fig.~\ref{fig:features}c. The separate treatment of paths and dilated pores explains the better prediction performance of CNN than the Kozeny--Carman equation. Specifically, the Kozeny--Carman equation views the porosity in dilated pores the same as that in channels and thus significantly overestimates the permeability. Such overestimation is absent in CNN predictions since the dilated pores are treated as a different category of fluid-filled porosity from the channels.

It should be noted that the features presented in Fig.~\ref{fig:features} are only for a qualitatively visualization of how CNN learned different patterns from the images. In practice, the parameters of CNN are determined by minimizing the prediction error of the training set with some techniques to prevent overfitting (e.g., using dropout or imposing sparsity). This is a more mathematically rigorous definition of the learning objective of CNN compared with analyzing the learned features. For instance, out of the ten filtered results from the first convolutional layer, the other eight features (not shown) have similarities compared to the two presented in Fig.~\ref{fig:features}. However, the number of features is still optimal in representing the information within the whole training dataset.

The images used in this study are synthetic and do not correspond to any real rocks. Hence, at this time we cannot project the performance of CNN with real rock geometries. The permeability studied in this work varied by only one order of magnitude, partly because of the unit employed to express permeability (pixel$^2$) and partly because of the synthetic nature of the geometries. Real porous media have permeability covering several orders of magnitudes. To use CNN to predict the permeability of different types of rocks, first, proper conversion of permeability is needed. For instance, images of sandstone are generally scanned with $\mu$m resolutions and those of shale are generally scanned with nanometer resolutions. Therefore, conversion from pixel-based permeability to dimensional permeability should in part differentiate these two kinds of rocks. Second, it is important that the geometries used for training are realistic. Additionally, different rock types, such as sandstones and shale, with their different pore structures, are likely to require their respective training data. This work mainly focuses on the proof-of-concept for adopting image recognition and specifically CNN/physics-informed CNN in predicting permeability of porous media. In order to predict on real rock samples, approaches that have been proven effective in other machine learning pratices~\cite{pan2010survey} can be adopted, such as adding some real rock images with known permeability into the training data, and maintaining a validation dataset of real rock images to further ensure the extrapolation capability when the majority of training samples are synthetic and the objectives are real rock images.

\section{Conclusion}
\label{sec:conclusion}
Fast predictions of physical properties of porous media are of significant practical importance. In this work, we propose a physics-informed convolutional neural network (CNN) to predict permeability from pore-scale images. The framework consists of the following components: (1) obtaining images of porous media, (2) building training datasets via fluid dynamics simulations, (3) training a physics-informed CNN, and (4) applying the trained model to predict new images that are not in the training set.  The predicative capability of the proposed model is demonstrated for synthetic images with a wide range of porosity and with various fractions of dilated pores as micro-scale heterogeneity. The predicted permeabilities for most samples have less than $10\%$ error compared to the lattice Boltzmann simulation results. In particular, for images with dilated pores where one cannot apply the Kozeny--Carman equation to estimate their permeabilities, the proposed model can give much better predictions. The CNN-based permeability prediction method is orders of magnitude faster than direct simulations using lattice Boltzmann. The proposed framework should have a great potential in geoscience and engineering applications and perhaps beyond. It can certainly be used to predict other physical properties of porous media as long as they are solely governed by the geometry.


\end{document}